\documentclass[journal,comsoc]{IEEEtran}
\usepackage{amsmath,amsfonts}
\usepackage{algorithmic}
\usepackage{algorithm}
\usepackage{array}
\usepackage[caption=false,font=normalsize,labelfont=sf,textfont=sf]{subfig}
\usepackage{textcomp}
\usepackage{stfloats}
\usepackage{url}
\usepackage{verbatim}
\usepackage{graphicx}
\usepackage{cite}
\usepackage{xcolor}
\def\fbinary{\mathbb{F}_2}

\def\symbol{\underline{\mathcal{X}}}
\def\symbolY{\underline{\mathcal{Y}}}
\def\lsymbol{\underline{x}}
\def\lsymbolY{\underline{y}}
\def\codebook{\mathcal{C}}
\def\baseset{\mathcal{B}}
\def\redset{\mathcal{I}}
\def\symcodeword{\mathcal{X}}
\def\symoutput{\mathcal{Y}}
\def\littlebitcodeword{\mathtt{x}}
\def\modulation{\mathcal{M}}

\DeclareMathOperator*{\argmax}{arg\,max}

\begin{document}
\title{Enhanced GCD through ORBGRAND-AI: Exploiting Partial and Total Correlation in Noise\\
{}
\thanks{This work was supported by the Defense Advanced Research Projects Agency (DARPA) under Grant HR00112120008, and by the National Science Foundation under Grant ECCS-2433994 and ECCS-2433996}
}

\author{Jiewei~Feng, Ken R. Duffy and Muriel M\'edard
\thanks{J. Feng and K. R. Duffy, Northeastern University, (e-mails: \{feng.ji,k.duffy\}@northeastern.edu).}
\thanks{M. M{\'e}dard, Massachusetts Institute of Technology, (e-mail: medard@mit.edu).}
}

\maketitle

\begin{abstract}
There have been significant advances in recent years in the development of forward error correction decoders that can decode codes of any structure, including practical realizations in synthesized circuits and taped out chips. While essentially all soft-decision decoders assume that bits have been impacted independently on the channel, for one of these new approaches it has been established that channel dependencies can be exploited to achieve superior decoding accuracy, resulting in Ordered Reliability Bits Guessing Random Additive Noise Decoding Approximate Independence (ORBGRAND-AI). Building on that capability, here we consider the integration of ORBGRAND-AI as a pattern generator for Guessing Codeword Decoding (GCD). We first establish that a direct approach delivers mildly degraded block error rate (BLER) but with reduced number of queried patterns when compared to ORBGRAND-AI. We then show that with a more nuanced approach it is possible to leverage total correlation to deliver an additional BLER improvement of ${\sim} 0.75$ dB while retaining reduced query numbers.  
\end{abstract}

\begin{IEEEkeywords}
Soft input, correlation, GRAND, GCD
\end{IEEEkeywords}

\section{Introduction}
\IEEEPARstart{C}{ode-agnostic} decoders have been the subject of considerable research attention, 
including Guessing Random Additive Noise Decoding (GRAND) \cite{8630851,solomon20, duffy2022_ordered,Chatzigeorgiou2023SymbolGRAND,9992258,chatzigeorgiou2024guessing,Allahkaram2025ConstrainedGRAND,Lukas2025SYGRAND}, Guessing Codeword Decoding (GCD) \cite{2024ZhengLCGCD,MA2025GCD,Wang2025ORBGCD}, and Ordered Reliability Direct Error Pattern Testing (ORDEPT)\cite{ORDEPT2023,ORDEPT2025journal}. These decoders offer the flexibility of decoding different codes with a single algorithmic architecture.
The practical promise of these approaches has been established primarily for GRAND variants, where its inherent parallelizability and the simplicity of its essential operations have resulted in synthesised circuits, e.g. \cite{Abbas2020HighThroughput_GRAND,Abbas2021ORBGRAND,Condo2021ORBGRAND, Abbas2021HighThroughput_GRAND_Markov,ji2025efficient,abbas2025improved}, and taped out chips, e.g. \cite{Riaz2021MultiCode,Burg2024GRANDAB,Riaz25ORBGRAND}, for both hard and soft detection systems.

One line of GRAND's development focuses on correlated channels, where variants have been introduced that exploit statistical knowledge of correlation to improve decoding, including GRAND Markov Order (GRAND-MO) \cite{An2022GRANDAI,zhan2022noise} for hard input and Ordered Reliability Bits GRAND Approximate Independence (ORBGRAND-AI) 
\cite{Duffy23ORBGRANDAI,orbgrandai2025Journal} 
for soft input.
The latter development leverages ideas from thermodynamic probability theory, e.g. \cite{lewis1995entropy}, to achieve multiple dB decoding improvement in channels with memory. ORBGRAND-AI's practicality has been demonstrated in a taped out chip \cite{orbgrandai2025Ece}.

A natural question arises as to whether the principles underlying ORBGRAND-AI can be adapted for use with GCD. Here we first demonstrate that a direct combination of ORBGRAND-AI with GCD results in mildly degraded BLER in return for much fewer queries. With a more nuanced understanding of how the theory of Approximate Independence is used in ORBGRAND-AI as well as the decoding process of GCD, we establish that this combination can be modified to leverage total channel state information (CSI), resulting in up to an additional ${\sim} 0.75$ dB improvement in BLER over ORBGRAND-AI while retaining some query reduction.  

The remainder of this paper is organized as follows. Section \ref{sec_prelim} introduces preliminaries, including summaries of ORBGRAND-AI and GCD. Section \ref{ORBAIGCD} describes the two proposed decoding algorithms whose performance is shown in Section \ref{sec_simulation}. Conclusions can be found in Section \ref{Sec_conclusion}.

\section{Preliminaries}\label{sec_prelim}
While GRAND decoders can decode non-linear codes, here we restrict our attention to binary linear codes as usually considered with GCD. A $k$-bit binary message $U^k=(U_1,\ldots,U_k)\in\fbinary^k$ is encoded with the $k\times n$ binary generator matrix $G$ resulting in $X^n=U^kG$. The codebook is denoted by the set $\codebook$. Let $H$ denote the codebook's parity check matrix. A sequence $X^n$ is a codeword if and only if $X^nH^T=0^{n-k}$, the zero vector of length $n-k$. The codeword $X^n$ is modulated to $\symcodeword^{n_s}=\modulation(X^n)$, where $n_s$ is the number of symbols needed to represent $n$ bits and $\modulation$ represents the modulation function. $\symcodeword^{n_s}$ is transmitted through a noisy channel and, after equalization, the channel output is $\symoutput^{n_s}=\symcodeword^{n_s}+\mathcal{N}^{n_s}$ where $\mathcal{N}^{n_s}$ represents the continuous noise effects impacting the codeword. A maximum likelihood (ML) decoder returns the codeword that achieves the maximum of $f_{\symoutput^{n_s}|X^n}(y^{n_s}|\littlebitcodeword^n)$ for $\littlebitcodeword^n\in\codebook$, where $f$ is the probability density function that characterizes the channel and can be calculated using CSI.

\textbf{ORBGRAND-AI:} ORBGRAND-AI \cite{orbgrandai2025Journal} originates from symbol-level ORBGRAND \cite{an2023soft}, which performs soft decoding at the symbol level without bit demapping. ORBGRAND-AI further leverages correlated channel effects to achieve enhanced decoding performance. The key idea underlying ORBGRAND-AI, known as approximate independence, can be understood as follows. The sequence $\symoutput^{n_s}$ is split into $n_b$ blocks with each block containing $b$ symbols, where we assume $b$ and $n_b=n_s/b$ are integers for notational simplicity:  $\symbolY^{n_b}=(\symbolY_1,\ldots,\symbolY_{n_b})$
\begin{align*}
    &=\left(\symoutput_1,\ldots ,\symoutput_{b},|\symoutput_{b+1},\ldots,\symoutput_{2b},|\ldots,|\symoutput_{n_s-b+1},\ldots,\symoutput_{n_s}\right).
\end{align*}
Similarly, every $b$ symbols in $\symcodeword^{n_s}$ are combined into one block, resulting in $\symbol^{n_b}=(\symbol_1,\ldots,\symbol_{n_b})$. Each block can also be seen as a higher-level symbol that contains multiple symbols. We denote $\lsymbol^{n_b}=\littlebitcodeword^n$ if the two sequences represent the same $n$-bit binary sequence. Each symbol is treated separately, and CSI is used to evaluate the PDF $f_{\symbolY_i|\symbol_i}(\lsymbolY_i|\lsymbol_i)$, which is denoted as $f(\lsymbolY_i|\lsymbol_i)$. The block-level hard demodulation for each block is given by $\lsymbol_i^*=\argmax_{\lsymbol_i}f(\lsymbolY_i|\lsymbol_i)$. 

By ignoring the correlation between blocks, the likelihood of a decision $\hat{\lsymbol}^{n_b}$ can be approximated as 
\begin{align}
    &f(\lsymbolY^{n_b}|\hat{\lsymbol}^{n_b})
    \approx\prod_{i=1}^{n_b}f(\lsymbolY_i|\hat{\lsymbol}_i)\label{eq_approx_likelihood}\propto \exp\left(\sum_{i=1}^{n_b}\log\frac{f(\lsymbolY_i|\hat{\lsymbol}_i)}{f(\lsymbolY_i|\lsymbol^*_i)}\right),
\end{align}
where we call
$\delta(\hat{\lsymbol}_i,\lsymbol^*_i)=-\log [f(\lsymbolY_i|\hat{\lsymbol}_i)/{f(\lsymbolY_i|\lsymbol^*_i)}]\geq 0$ the \textit{relative reliability} of the demodulated block $\lsymbol_i^*$ to the alternative block $\hat{\lsymbol}_i$. The relative reliability reduces to bit-level reliability, i.e. the absolute value of a log-likelihood ratio, in the absence of any correlation. To rank order the likelihood of decisions from most likely to least likely, it suffices to rank order the sum of relative reliabilities $\sum_{i=1}^{n_b}\delta(\hat{\lsymbol}_i,\lsymbol_i^*)$ from least to highest. Each term $\delta(\hat{\lsymbol}_i,\lsymbol_i^*)$ quantifies the degradation of likelihood resulting from replacing the hard demodulated block $\lsymbol_i^*$ with $\hat{\lsymbol}_i$.

For the $i$-th block, there are $|\symcodeword|^{b}$ possible decisions of $\symbol_i$ with $\symbol_i^*$ having the least relative reliability, of zero. This results in $n_b(|\symcodeword|^{b}-1)$ positive relative reliabilities for the alternative decisions except for the hard demodulated blocks. When these positive relative reliabilities are sorted and plotted against their rank, they fall approximately along a straight line. This structural regularity enables the generator from ORBGRAND \cite{duffy2022_ordered} to be adapted for pattern generation in correlated channels. At each step, an $n_b$ by $(|\symcodeword|^{b}-1)$ binary matrix $Q$ is generated to indicate which alternative decisions of the blocks are desired. The entry $Q_{i,j}=1$ indicates \textit{swapping} the hard demodulation $\lsymbol^*_i$ of the $i$-th block with the $j$-th alternative decision for the same block. If all entries of the $i$-th row are zero, then we retain the hard demodulation $\lsymbol^*_i$ of the $i$-th block. Each hard demodulated block can be swapped at most once, hence a generated $Q$ is valid if each row has at most one non-zero entry. Each valid $Q$ corresponds to a unique decision for the blocks, $\hat{\lsymbol}^{n_b}$, while an invalid $Q$ is discarded. Since the decisions are generated in approximately decreasing order of likelihood, the first decision $\hat{\lsymbol}^{n_b}$ that passes the parity check is a near-ML decoding.

As demonstrated in \cite{orbgrandai2025Journal}, ORBGRAND-AI offers superior decoding accuracy by leveraging CSI for correlated channels
and this improvement can be understood through \eqref{eq_approx_likelihood} using the lens of Information Theory. Suppose the noise effects $\mathcal{N}^{n_s}$ follow a zero-mean multivariate-Gaussian distribution with variance $\sigma^2$, where two noise effects $\mathcal{N}_i$ and $\mathcal{N}_j$ within the same block have covariance $\text{Var}[\mathcal{N}_i,\mathcal{N}_j]=\sigma^2\rho^{|i-j|}$ and $\text{Var}[\mathcal{N}_i,\mathcal{N}_j]=0$ if $i,j$ are not in the same block. Using equation (9.34) in \cite{cover1991elements}, the normalized differential entropy (NDE) is given by $h(\mathcal{N}^{n_s}|b)=\log(2e\pi)+\left(1-\frac{1}{b}\right)\log(1-\rho^2)$.
When $b=n_s$, this corresponds to the true CSI where all noise effects are correlated and $h(\mathcal{N}^{n_s}|n_s)=\log(2e\pi)+(1-\frac{1}{n_s})\log(1-\rho^2)$. When $b=1$, this corresponds to an Additive White Gaussian Noise (AWGN) channel where noise effects are independent and $h(\mathcal{N}^{n_s}|1)=\log(2e\pi)>h(\mathcal{N}^{n_s}|n_s)$. This difference indicates the degradation of performance when using AWGN model to decode correlated Gaussian channels. More generally, this difference illustrates the degradation resulting from using a decoder that assumes memoryless channels and interleavers. By considering the right hand side of \eqref{eq_approx_likelihood} for approximation, the NDE of the assumed model in decoding approaches the true NDE for $b \rightarrow n_s$. Moreover, by setting $b=2$, we can recover more than half of the difference in NDE between a memoryless channel and the true CSI while having the ability to decode efficiently in practice \cite{orbgrandai2025Ece}.

\textbf{GCD:} GCD \cite{MA2025GCD} assumes a binary-discrete memoryless channel, in which case the channel output can instead be represented by $Y^n$. As a result, the likelihood of $\littlebitcodeword^n$, denoted as $f_{Y^n|X^n}(\mathtt{y}^n|\littlebitcodeword^n)$, can be calculated as $\prod_{i=1}^{n} f_{Y_i|X_i}(\mathtt{y}_i|\littlebitcodeword_i)$. The bit-level hard demodulation is given by $\littlebitcodeword_i^*=\argmax_{\littlebitcodeword_i} f_{Y_i|X_i}(\mathtt{y}_i|\littlebitcodeword_i)$. Without loss of generality, GCD assumes that the parity check matrix is written in a systematic format, i.e., $H=[I_{n-k}|B]$ where $B$ is an $n-k$ by $k$ binary matrix and $I_{n-k}$ is an $n-k$ by $n-k$ identity matrix. This also implies that $G$ can be written as $[ B^T|I_k]$. This assumption will be referred to as the \textit{systematic assumption} and can be achieved for any binary linear code by using Gauss-Jordan elimination and column permutation, which only needs to be performed once offline. For GCD, let $\redset=\{1,\ldots,n-k\}$ and $\baseset=\{n-k+1,\ldots,n\}$. Let $x_{\baseset}$ denote the subsequence of $x$ containing the entries with indices in $\baseset$. This systematic assumption enables GCD to extend any binary subsequence $\hat{\littlebitcodeword}_{\baseset}$ to a complete codeword $\hat{\littlebitcodeword}^n$ via $\hat{\littlebitcodeword}_{\redset}=\hat{\littlebitcodeword}_{\baseset}B^T$. We will call the bits in $\hat{\littlebitcodeword}_\baseset$ the \textit{base bits} as they are extrapolated to identify a whole codeword. 

In GCD, a pattern generator is used to sequentially provide decisions of the base bits. For a decision of the base bits denoted as $\hat{\littlebitcodeword}^n_{\baseset}$, its likelihood is calculated as $\prod_{i\in\baseset} f_{Y_i|X_i} (\mathtt{y}_i|\hat{\littlebitcodeword}_i)$, where the extended codeword has likelihood $f_{Y^n|X^n}(\mathtt{y}^n|\hat{\littlebitcodeword}^n)$. GCD keeps a record of the running maximum of $f_{Y^n|X^n}(\mathtt{y}^n|\hat{\littlebitcodeword}^n)$, which is denoted as $p^*$, and the corresponding codeword. Assuming the pattern generator generates decisions for the base bits in decreasing order of $\prod_{i\in\baseset} f_{Y_i|X_i}(\mathtt{y}_i|\hat{\littlebitcodeword}_i)$, GCD terminates when 
\begin{align}
    \prod_{i\in\baseset} f_{Y_i|X_i}(\mathtt{y}_i|\hat{\littlebitcodeword}_i)\prod_{i\in\redset} f_{Y_i|X_i}(\mathtt{y}_i|\littlebitcodeword_i^*)<p^*,\label{eq_stopping_GCD}
\end{align} 
as no future pattern could lead to a codeword that has higher likelihood than the running maximum $p^*$. The codeword whose likelihood is the running maximum is the ML decoding. If the decisions of base bits are generated in approximately decreasing order of likelihood, the resulting decoding is a near-ML decoding.

\section{GCD driven by ORBGRAND-AI}\label{ORBAIGCD}
We investigate the use of ORBGRAND-AI as the pattern generator of GCD to incorporate correlations. When leveraging correlations instead of assuming independent noise effects, column permutation to place $H$ in systematic format would require modifying the order of transmission of bits at the transmitter side. Following MATLAB's notation, let $H_{:,A}$ denote the submatrix of $H$ containing the columns with indices in the integer set $A$. To integrate ORBGRAND-AI with GCD for all linear binary codes, including all systematic and non-systematic codes, we assume Gauss-Jordan elimination without column permutation results in 
\begin{align}
    H_{:,\baseset}=B, H_{:,\redset}=I, \label{eq_semi_sys_H}
\end{align}
for some $\baseset\subseteq\{1,\ldots,n\}$ with $|\baseset|=k$ and $\redset=\{1,\ldots,n\}\setminus\baseset$. Equivalently, the set $\redset$ contains the column indices in $H$ that form an identity matrix, while $\baseset$ contains the rest of the column indices that form the matrix $B$. Note that, in this format, any subsequence $\hat{\littlebitcodeword}_{\baseset}$ can be extended to a complete codeword via  $\hat{\littlebitcodeword}_{\redset}=\hat{\littlebitcodeword}_{\baseset}B^T$. Equation (\ref{eq_semi_sys_H}) differs from the systematic assumption in the sense that $\baseset$ and $\redset$ can be non-consecutive integer sets. The non-consecutiveness can occur for some non-systematic codes, as illustrated in Section \ref{sec_simulation}. 

As in ORBGRAND-AI, consecutive symbols are grouped into $n_b$ blocks, with each block containing $b$ symbols or fewer if needed. In addition, we assume that each block only contains either bits in $\baseset$ or bits in $\redset$. This ensures that the $n_b$ blocks are separated into blocks containing only the base bits and blocks containing the bits in $\redset$. The separation allows us to use ORBGRAND-AI to generate decisions for the base bits and extend the decisions using $B$. An exception occurs when a symbol contains both base bits and bits in $\redset$. In that case, we adopt the framework of Locally Constrained GCD\cite{2024ZhengLCGCD} (LC-GCD) as follows: we use ORBGRAND-AI to generate decisions for the symbols that contain at least one base bit. Even though the generated partial decisions will be overspecified, each of them will be examined by the parity check matrix to see if it can lead to a valid codeword. A partial decision that passes the parity check will be extended using the decision of the base bits, as in GCD.

Let $\{\symbol_a\}_{a\in\baseset'}$ denote the \textit{base blocks} containing only the base bits, where $\baseset'\subseteq\{1,\ldots,n_b\}$ contains indices of blocks. Similarly, let $\{\symbol_a\}_{a\in\redset'}$ denote the rest of the blocks.
Every specification of the base blocks $\{\symbol_{a,i}\}_{a\in\baseset'}$ can be extended to a full codeword using $H$ similarly to GCD. Note that we only group symbols with consecutive indices into blocks, as noise effects on consecutive symbols are more correlated than the noise effects on separated symbols. For example, symbols $\symbol_3,\symbol_4,\symbol_5$ can be grouped into a block but $\symbol_3,\symbol_4,\symbol_8$ will not be grouped into a block. To illustrate the reason, assume a Gauss-Markov channel with correlation $\rho$ and BPSK. The noises $N_i$ and $N_j$ have correlation $\rho^{|i-j|}$ which decreases exponentially as $|i-j|$ increases. Therefore, when $\baseset$ or $\redset$ contains non-consecutive integers, not every group of $b$ consecutive symbols is combined into one block. Further illustrations will be given in Section \ref{sec_simulation} using an example. 

\subsection{ORBGRAND-AI driven GCD (direct combination)}
With ORBGRAND-AI, we can sequentially generate decisions of the base blocks and use \eqref{eq_approx_likelihood} to calculate the likelihoods of decisions. This combination is summarized as follows:
1. \textbf{Generator:} ORBGRAND-AI is used to generate partial decisions of base blocks denoted as $\hat{\symbol}_{\baseset',i}=\{\hat{\symbol}_{a,i}\}_{a\in\baseset'}$ for the $i$-th query. Each partial decision is then extended through $\hat{\symbol}_{\redset',i}=\hat{\symbol}_{\baseset,i}B^T$ in the binary field.\\
2. \textbf{Maximum update condition:} The likelihood of each generated decision is calculated by $\prod_{a=1}^{n_b}f(\symbolY_a|\hat{\symbol}_{a,i})$. For a running maximum likelihood $p^*$, it is updated to be the likelihood of the $i$-th query if 
\begin{align}
    \prod_{a=1}^{n_b}f(\symbolY_a|\hat{\symbol}_{a,i})>p^*.\label{eq_update_rule_partial}
\end{align}
3. \textbf{Stopping criterion}: Inspired by \eqref{eq_stopping_GCD}, the generator keeps providing queries until the following is met:
\begin{align}    
\left(\prod_{a\in\baseset'}f(\symbolY_a|\hat{\symbol}_{a,i})\right)\left(\prod_{a\in\redset'}f(\symbolY_a|\symbol_{a}^*)\right)<p^*.\label{eq_stopping_partial}
\end{align}
This stopping criterion is consistent with the pattern generator, as both incorporate the correlation within each block while assuming independence between blocks.

We will refer to this combination as the \textit{direct combination} where \eqref{eq_update_rule_partial} and \eqref{eq_stopping_partial} use partial CSI through \eqref{eq_approx_likelihood}. 

\subsection{ORBGRAND-AI integrated GCD (advanced combination)}
Both ORBGRAND-AI and the direct combination use $(\ref{eq_approx_likelihood})$ that leverages partial CSI throughout their operations. This indicates potential for further improvement based on the difference between $h(\mathcal{N}^{n_s}|b)$ and $h(\mathcal{N}^{n_s}|n_s)$ when $b<n_s$. Note that the assumption of independence between blocks is only used for pattern generation, while we utilize that omitted correlation into the maximum update rule. For the $i$-th generated decision $\lsymbol^{n_b,i}=(\lsymbol_{1,i},\ldots,\lsymbol_{n_b,i})$, the running maximum $p^*$ is updated if \begin{align}
    f(\lsymbolY^{n_b}|\lsymbol^{n_b,i})>p^*,\label{eq_update_total} 
\end{align}
where $f(\lsymbolY^{n_b}|\lsymbol^{n_b,i})$ is evaluated directly using the full CSI instead of using the approximation through \eqref{eq_approx_likelihood}. 

For the stopping criterion, note that the second product in (\ref{eq_stopping_partial}), which serves as an upper bound for the likelihood multiplier corresponding to the blocks in $\redset'$, relies on block-level hard demodulation and the assumption of independence between blocks. The principle here is to use the approximate independence to facilitate efficient pattern generation, but to keep full CSI to inform the maximum update condition. 

We will refer to the direct combination with the maximum update condition replaced by (\ref{eq_update_total}), which leverages total CSI, as advanced combination. 
The replacement of (\ref{eq_update_rule_partial}) by (\ref{eq_update_total}) can be seen as changing the block size $b$ to $n_s$ in the calculation of block-level likelihood. 

\section{Performance Evaluation}
\label{sec_simulation}
Given a set of autocovariance constraints, the minimal-order Gauss-Markov process consistent with those constraints  achieves the highest entropy among all stationary processes satisfying those constraints \cite{Choi}. We consider the first order Gauss-Markov process which arises, for example, in an equalized inter-symbol-interference (ISI) channel subject to AWGN. Let $Z_1,\ldots,Z_n$ be independent and identically distributed Gaussian random variables with mean zero and variance $\sigma^2$. The definition of the Gauss-Markov process for BPSK is given as follows, whose generalization to higher order modulation will become clear. A Gauss-Markov process, $\{N^n\}$, with correlation $\rho$ is defined by $N_1=Z_1$ and $N_i=\rho N_{i-1}+\sqrt{1-\rho^2}Z_i$ for $i=2,\ldots,n$. The received signal is $Y^n=(1-2X^n)+N^n\in \mathbb{R}^n$.
For any $m\leq n$, we have that 
\begin{align}       
    f_{Z^m}(z^m)=f_N(z_1)\prod_{i=2}^m f_N\left(\frac{z_i-\rho z_{i-1}}{\sqrt{1-\rho^2}}\right),\label{eq_density}
\end{align}
where $f_N$ is the PDF of the mean $0$ Gaussian distribution with variance $\sigma$.
The likelihood of a decision $\hat{x}^m$ needed in the decoding process is calculated by
$f_{Y^m|X^m}(\mathtt{y}^m|\hat{\littlebitcodeword}^m)=f_{Z^m}(\mathtt{y}^m- (1-2\hat{\littlebitcodeword}^m)).$ For higher order modulation, $N_i$ is then replaced by $\mathcal{N}_i=(N_i,N'_i)$ to represent the noise effects in complex plane, where $\{N_i\}_{i=1}^{n_s}$ and $\{N'_i\}_{i=1}^{n_s}$ are two independent Gauss-Markov processes defined above. In both cases, \eqref{eq_density} implies that the complexity of evaluating the likelihood of a block decision with length $b$ is $O(b)$, and hence both \eqref{eq_update_rule_partial} and \eqref{eq_update_total} have the same complexity $O(n_s)$ for each pattern.

\begin{figure}
    \centering
    \includegraphics[width=0.9\linewidth]{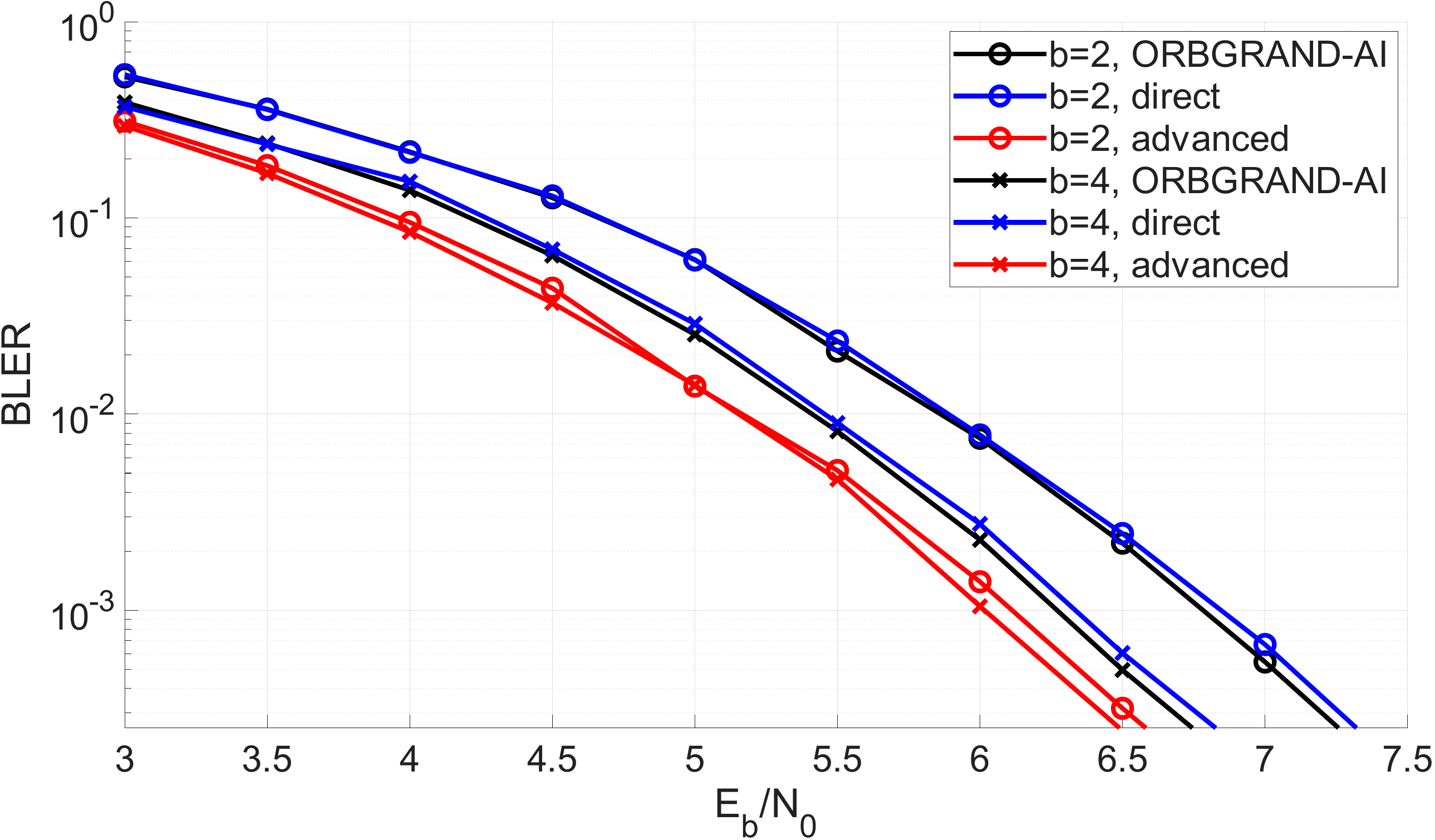}
    
    \caption{BLER vs ${E_b/N_0}$ for CRC $[64,48]$ using BPSK and $\rho=0.5$.}
\label{GRANDAI_CRC_64_48_1}
\end{figure}

We first investigate the two combinations when the code is systematic. Fig. \ref{GRANDAI_CRC_64_48_1} records BLER using CRC $[64,48]$, where the $16$-bit CRC is specified in the Distributed Network Protocol. CRC codes are widely used for error detection, but GRAND and GCD can upgrade them to error correction. We set $\rho=0.5$ to represent reasonable levels of ISI. For systematic codes where the first $k$ bits are information bits, the set $\baseset$ hence contains consecutive integers. ORBGRAND-AI has mildly superior performance to direct combination in terms of BLER. In contrast, advanced combination, which uses the full CSI in its update rule, experiences a ${\sim} 0.75$ dB gain when all decoders use $b=2$ and around ${\sim}0.4$ dB when all decoders have $b=4$. 

\begin{figure}
    \centering
    \includegraphics[width=0.9\linewidth]{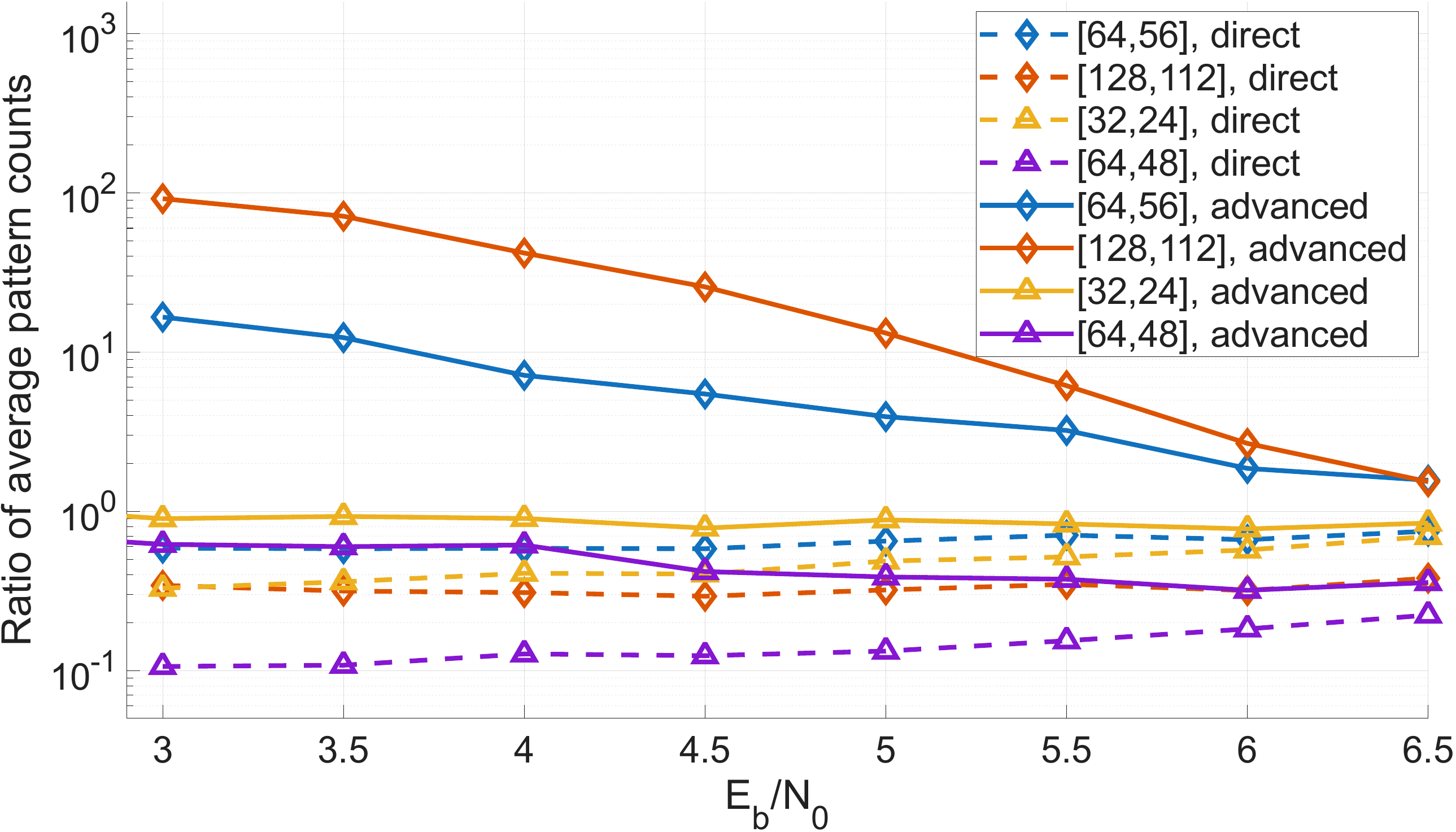}
    \caption{Ratios of average pattern counts to ORBGRAND-AI for $b=2$.}
    \label{pattern_ratio_CRC_rho}
\end{figure}

Fig. \ref{pattern_ratio_CRC_rho} further investigates the reduction of pattern counts for different code dimensions using BPSK and $b=2$. The dashed lines represent the ratio of average pattern counts between direct combination and ORBGRAND-AI, while the solid lines represent the ratio between advanced combination and ORBGRAND-AI. Direct combination demonstrates consistent query reduction for all code rates and code lengths, where the reduction is more significant for codes with lower rate when $n$ is fixed. As a counterpoint, to obtain better BLER, advanced combination has less query reduction gain for codes with a lower rate and can require more queries compared to ORBGRAND-AI for codes with a higher rate.

\begin{figure}
    \centering
    \includegraphics[width=0.9\linewidth]{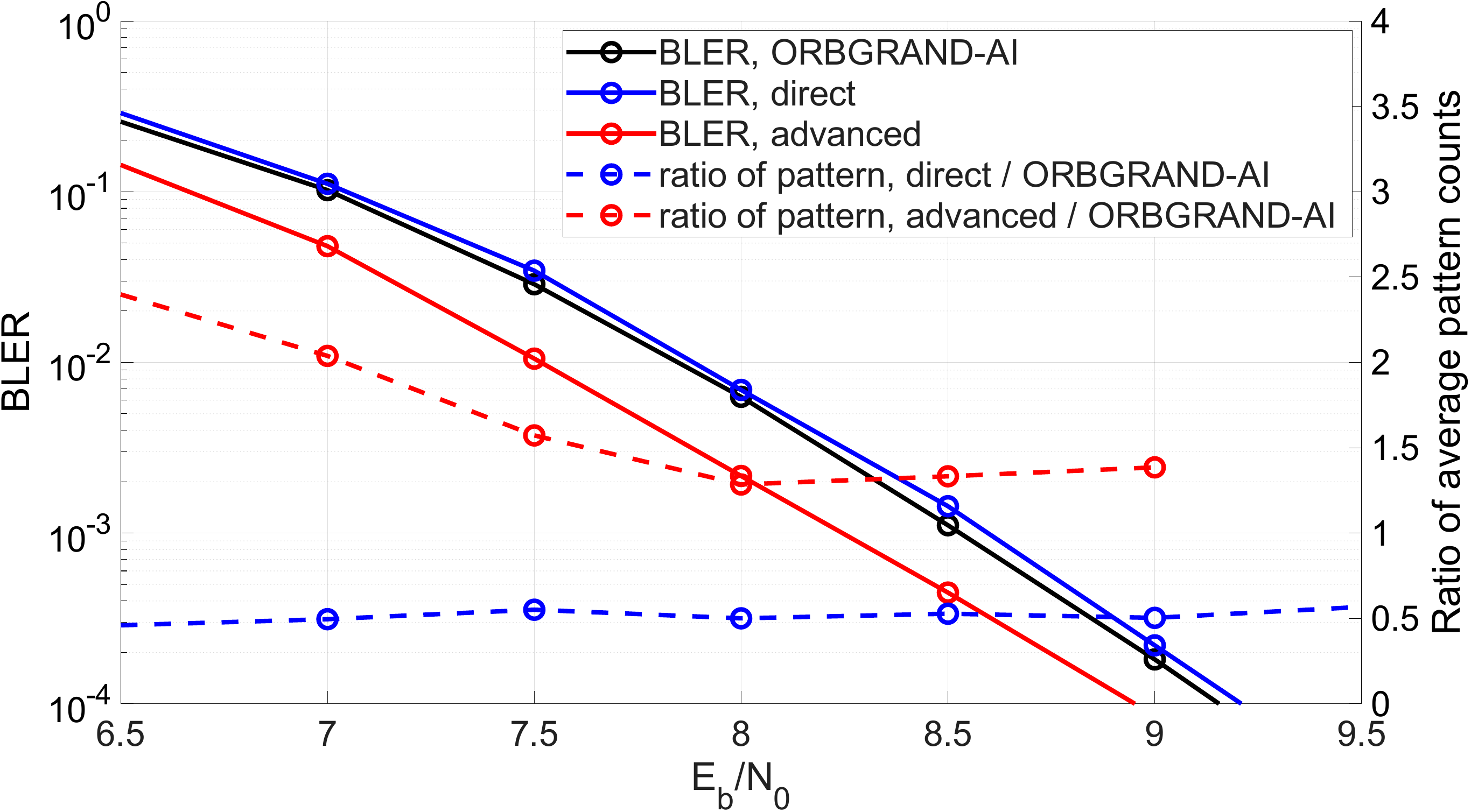}
    
    \caption{BLER and ratio of pattern count for CRC $[256,238]$. 16-QAM subject to Gauss-Markov noise with $\rho=0.5$. Decoders operate with $b=2$. }
\label{CombineBLERandQueryRatio_CRC_256_238_1_rho}
\end{figure}

Fig. \ref{CombineBLERandQueryRatio_CRC_256_238_1_rho} records BLER (left y-axis) and query count ratio (right y-axis) under 16-QAM. In each decoding, all decoders have maximum number of queries set to $10^6$. For demonstration, we use CRC [256,238] code where there exists one symbol containing two base bits and two bits in $\redset$.
In this case, we adopt LC-GCD as mentioned. As depicted by the figure, similar improvement can be obtained compared to Fig. \ref{pattern_ratio_CRC_rho} in terms of query reduction. Note that the BLER improvement using advanced combination is less significant but is expected to increase for larger values of $n$. This is because in higher order modulation, correlation occurs in two ways: 1, correlation within each symbol, which is already sufficiently leveraged by the symbol level generation in ORBGRAND-AI. 2, ISI that advanced combination leverages to improve decoding.

\begin{figure}
    \centering
    \includegraphics[width=0.9\linewidth]{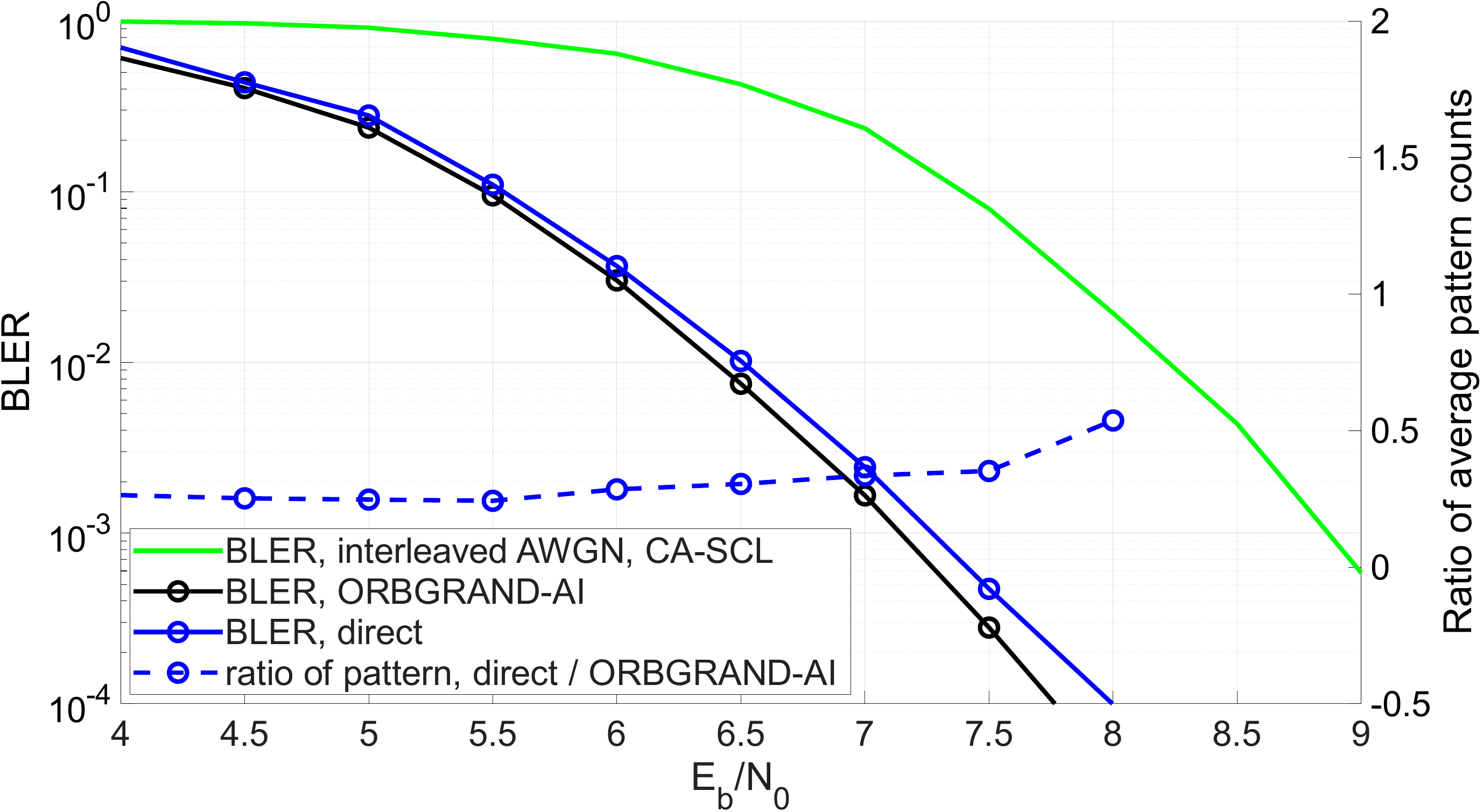}
    
    \caption{BLER and ratio of pattern count for a CA-Polar $[128,110]$ code.}
\label{GRANDAI_CAPOLAR_128_110_1_rho}
\end{figure}

When decoding non-systematic codes, the sets $\baseset$ and $\redset$ can both contain consecutive integers. In this case, the blocks are constructed as if the code is systematic and therefore the decoding is performed as previously illustrated. If this is not the case, however, the set of base bits $\baseset$ and its complement $\redset$ no longer contain only consecutive numbers, which makes one or more blocks contain fewer than $b$ bits. We illustrate the resolution of this issue using a $[128,110]$ 5G NR CA-Polar code with an $11$-bit CRC and BPSK. The base bits of this code are $\baseset=\{18,\ldots,32,34,\ldots,128\}$ and the complement is $\redset=\{1,\ldots,17,33\}$. For $b=2$, every two bits in $\redset$ can be grouped into one block except for the $17$-th bit and the $33$-th bit as they are non-consecutive. When bits $17$ and $33$ are transmitted consecutively, their experience of the channel is highly correlated, enabling this correlation to be leveraged for decoding. However, implementing such consecutive transmission would necessitate column permutation of $H$ and $G$, which requires modifications at the transmitter. By default, these two bits are transmitted with a time gap, which reduces their noise correlation to a near-negligible level, and hence can be ignored in the pattern generator as described in Section \ref{ORBAIGCD}. In this default configuration, each of the two bits becomes an individual block when using direct combination, resulting in a major reduction in average query number at the cost of a minor BLER degradation, as depicted in Fig. \ref{GRANDAI_CAPOLAR_128_110_1_rho}. In addition, the green line represents CRC-aided Successive Cancellation List (CA-SCL) decoding with list size eight under interleaved AWGN as a benchmark. 

\section{Conclusions}\label{Sec_conclusion}

The direct combination of ORBGRAND-AI and GCD provides a reduction in average pattern counts. In contrast, the advanced combination provides better BLER performance. We note that techniques have been developed to leverage binary linear code structure to reduce query numbers \cite{Rowshan22, Rowshan23, rowshan2023segmented,2024WangPCGRAND, Lukas2025Tree}. However, it remains unclear how those skipping techniques can be directly adapted to ORBGRAND-AI owing to its non-binary sensibilities.

\bibliographystyle{IEEEtran}

\bibliography{b.bib}

\end{document}